# Dark energy and quantum entanglement

Mark Ya. Azbel'

School of Physics and Astronomy, Tel-Aviv University,

Ramat Aviv, 69978 Tel Aviv, Israel[+];

Max-Planck-Institute für Festkorperforschung – CNRS,

F38042 Grenoble Cedex 9, France

**Abstract**

Entangled states in the universe may change interpretation of observations and even revise the concept of dark energy.

Recent observations of thermonuclear supernovae show that in the last 5 billion years the expansion of the Universe has not been steady or decelerating (as expected), but speeding up [1]. When combined with evidence from galaxy clustering, from the fluctuations in the cosmic microwave background, and bending of light by gravitational lenses, this points to a Universe that is 2/3 "dark energy" [2, 3], and whose expansion currently switches from accelerating to slowing down. Several candidates for the nature of the black energy (but not for current timing of the crossover to slowing down) were suggested, including a modern version of Einstein's cosmological constant [4]. However, estimates of dark energy contribution and even its concept may be altered by taking into account quantum coherent states in early universe. Strong many-body interactions in the dense early universe destroy their coherence, while rapid universe expansion separates coherent states by huge distances (which do not allow these states with only a fraction of particle density in each of them to "reunite" thereafter). The universe expansion changes its density, dominant particles, their states and interactions, whose knowledge is hardly comprehensive. However, if ultimately the expansion leaves some of the coherent states entangled [5], then a light beam, emitted by a distant source, travels to the Earth observer along several geodesic paths (see later). Its largest deviation (which may significantly alter the distance traveled by the beam) is related to the total number of particles in coherent states, while its intensity (thus its observed brightness) is proportional to the number of such particles interacting with the beam. One of the geodesics is not curved at all, intensity (thus observed brightness) of its beam is proportional to the fraction of coherent states, which are remote from and do not interact with the beam. This may revise conventional non-quantum interpretation of the beam bending, calculated brightness and the number of the beam sources, thus all such estimates of dark energy in the universe. Of course, accurate calculation, and even estimate of the fraction of superimposed states in current universe, poses formidable problems in quantum gravitation. Dephasing remains an unsettled problem even in much simpler case of well known mesoscopic systems [6], extensively studied experimentally and theoretically. However, one may start with an attempt at a refined interpretation of existing observations, complemented with theoretical, experimental and numerical study of known systems. Elucidate its main qualitative result with a simple non-relativistic quantum model.

Suppose quantum particles are located in four remote regions with no interaction between them (i.e. potential energy is assumed to be zero beyond certain distance) at the initial moment t=0. Thereafter particles from the "source" region S (their coordinates are $r_s$ ) move in the direction of the "lens" region L. After certain time they come sufficiently close to its, and only its, particles and interact with them. Particles in any region but S are localized in this region by its infinite potential walls. The states in the region L and (remote from it) region R are correlated; coordinates of all their particles $r_c$ are restricted either to L or to R. The fourth (distant from L and R) region is D, coordinates of its particles are $r_d$. The wave function $\psi$ yields the Schroedinger equation

$$i\hbar \partial \psi / \partial t = [\hat{H}_D(r_d) + \hat{H}_{LS}(r_c, r_s) + \hat{H}_R(r_c)]\psi \qquad (1)$$

where $\hat{H}$ is the Hamiltonian operator in the region denoted by the subscript (e.g. $\hat{H}_{LS}(r_c, r_s)$ accounts for the interaction of $r_c$ and $r_s$ which emerges with time). Present solution to Eq. (1) in the form

$$\psi = \psi_D(r_d; t) [a\psi_{LS}(r_c, r_s; t) + b\psi_R(r_c; t)\psi_S(r_s; t)] \qquad (2)$$

Such presentation is possible since in the considered model $\psi_{LS} = 0$ when $r_c$ is outside L and $\psi_R = 0$ when $r_c$ is outside R. By Eqs. (1) and (2),

$$i\hbar \partial \psi_D/\partial t = \hat{H}_D \psi_D; \quad i\hbar \partial \psi_{LS}/\partial t = \hat{H}_{LS}\psi_{LS}; \quad i\hbar \partial \psi_R/\partial t = \hat{H}_R \psi_R \qquad (3)$$

Initial conditions in the model are

$$\psi_D(r_d; 0) = \phi_D(r_d); \quad \psi_S(r_s; 0) = \phi_S(r_s);$$

$$\psi_{LS}(r_c, r_s; 0) = \phi_L(r_c)\phi_S(r_s); \quad \psi_R(r_c; 0) = \phi_R(r_c) \qquad (4a)$$

All functions $\phi$ are normalized, and

$$|a|^2 + |b|^2 = 1 \qquad (4b)$$

So,

$$\int |\psi|^2 dr_d\, dr_c = |a|^2 \int |\psi_{LS}(r_c, r_s; t)|^2 dr_c + |b|^2 |\psi_S(r_s; t)|^2 \qquad (5)$$

The second term in Eq. (5) does not depend on interaction with any other region, thus the second beam does not deviate. Its intensity (thus its observed brightness) is proportional to the fraction of coherent particles in the remote region. The first term deviates. Its intensity (thus its brightness) is proportional to the number of coherent particles in the lens region. The interaction of the beam with coherent lens particles is related in the Schroedinger Eq. (1) to characteristics (e.g. charge) of entire particles, it is independent of their fraction in the lens region. So, the deviation, thus the distance traveled by the beam to the observer, is related to the total number (thus total charge, or mass, etc) of coherent particles. The calculation is readily generalized to any number of regions with any number of coherent and non-coherent states in each region. It is a much more difficult problem how and in what time quantum entanglement is eliminated.

To summarize, scattering by entangled states, thus the calculated brightness of, geodesic distances to, and even number of beam sources, are different from those in non-quantum case, and may change the estimates for dark energy and possibly even its concept.